\documentclass[twocolumn,superscriptaddress,nopacs,aps,prx,reprint]{revtex4-2}
\usepackage{graphicx}
\usepackage{bm}
\usepackage{color}
\usepackage{xcolor}
\usepackage{amsfonts}
\usepackage[normalem]{ulem}
\usepackage{subfigure}
\usepackage{siunitx}
\DeclareSIUnit{\angstrom}{\textup{\AA}}
\usepackage{mathtools}
\renewcommand{\vec}[1]{\mathbf{#1}}

\newcommand{\di}{\mathrm{d}}
\newcommand{\kB}{k_\mathrm{B}}
\newcommand{\lambdaD}{\lambda_\mathrm{D}}
\newcommand{\Reff}{R_\mathrm{eff}}
\newcommand{\caF}{\mathcal{F}}
\newcommand{\caI}{\mathcal{I}}
\newcommand{\caM}{\mathcal{M}}
\newcommand{\caR}{\mathcal{R}}
\newcommand{\caT}{\mathcal{T}}
\newcommand{\TM}{\mathrm{\scriptstyle TM}}

\begin{document}

\title{Universal Casimir attraction between filaments at the cell scale}

\author{Benjamin Spreng}
\affiliation{Department of Electrical and Computer Engineering, University of California, Davis, CA 95616, USA}

\author{Hélène Berthoumieux}
\affiliation{Gulliver, CNRS, {\'E}cole Sup{\'e}rieure de Physique et Chimie Industrielles de Paris, Paris Sciences et Lettres Research University, Paris 75005, France}
\affiliation{Fachbereich Physik, Freie Universität Berlin, Arnimallee 14, Berlin, 14195, Germany}

\author{Astrid Lambrecht}
\affiliation{Forschungszentrum Jülich, 52425 Jülich, Germany}\
\affiliation{RWTH Aachen University, 52062 Aachen, Germany}

\author{Anne-Florence Bitbol}
\email{anne-florence.bitbol@epfl.ch }
\affiliation{
Institute of Bioengineering, School of Life Sciences, École Polytechnique Fédérale de Lausanne (EPFL), CH-1015 Lausanne, Switzerland}
\affiliation{SIB Swiss Institute of Bioinformatics, CH-1015 Lausanne, Switzerland}

\author{Paulo A. Maia Neto}
\email{pamn@if.ufrj.br}
\affiliation{Instituto de F\'{\i}sica, Universidade Federal do Rio de Janeiro \\ Caixa Postal 68528,   Rio de Janeiro,  RJ, 21941-972, Brazil}

\author{Serge Reynaud}
\email{serge.reynaud@lkb.upmc.fr}
\affiliation{Laboratoire Kastler Brossel, Sorbonne Universit\'e, CNRS, ENS-PSL, Coll\`ege de France, Campus Jussieu, F-75005 Paris, France }

\date{\today}

\begin{abstract}
The electromagnetic Casimir interaction between dielectric objects immersed in salted water includes a universal contribution that is not screened by the solvent and therefore long-ranged. Here, we study the geometry of two parallel dielectric cylinders. We derive the Casimir free energy by using the scattering method. We show that its magnitude largely exceeds the thermal energy scale for a large parameter range. This includes length scales relevant for actin filaments and microtubules in cells. We show that the Casimir free energy is a universal function of the geometry, independent of the dielectric response functions of the cylinders, at all distances of biological interest. While multiple interactions exist between filaments in cells, this universal attractive interaction should have an important role in the cohesion of bundles of parallel filaments. 
\end{abstract}

\maketitle

\section{Introduction}

The electromagnetic Casimir or van der Waals attraction between dielectric particles immersed in salted water~\cite{Mitchell1974,MahantyNinham1976,Israelachvili2011,Parsegian2006} was recently shown to be stronger and of longer range than previously expected. 
This long-range Casimir interaction was predicted as an effect of non-screened electromagnetic thermal fluctuations confined between plane dielectric surfaces \cite{MaiaNeto2019}. For spherical particles, the interaction is a universal function of distance, independent of the dielectric response functions of the particles~\cite{Schoger2022}, and it overtakes non-universal contributions at distances of the order of $0.1\,\mu$m.
On the other hand, such non-universal contributions 
dominate the total interaction when probing the force between dielectric spheres at distances in the nanometer range~\cite{Elzbieciak-Wodka2014,Smith2019,Smith2020}. 
The existence of the non-screened universal Casimir force was proven experimentally on a microsphere held by optical tweezers interacting with a larger rigidly held sphere at distances above $0.2\,\mu$m~\cite{Pires2021}. A long-ranged attraction at similar distances was also found for optically trapped dielectric microspheres in salted water~\cite{Hansen2005,Kundu2019}. 

Between spherical particles, this universal Casimir interaction only dominates the thermal energy scale $\kB T$ associated to Brownian motion in the liquid when the distance between spheres is smaller than one tenth of the smallest radius. 
However, there are other highly relevant geometries where this interaction should be more significant. Here, we study the case of two parallel dielectric cylinders in salted water, where the force is expected to be proportional to the length of the cylinders, itself much larger than the radial dimensions.
We show that the electromagnetic Casimir attraction in such configuration can indeed dominate the thermal energy $\kB T$ at distances larger than the radii.

Considering dielectric cylinders immersed in salted water allows us to address the following question: Can the universal Casimir interaction play an important role in biological systems at the cell scale? Indeed, filamentous structures are ubiquitous in cells. Cytoskeletal filaments, in particular actin filaments and microtubules, play crucial parts in maintaining the integrity of eukaryotic cell shape, in its deformations, as well as in multiple sub-cellular processes, by actively generating forces with the help of motor proteins~\cite{Salbreux12,Murrell15,Burla19}. Actin filaments form bundles, where filaments are cross-linked by specific proteins into parallel arrays. Microtubules, which are thicker and more rigid than actin filaments, also form bundles cross-linked by microtubule-associated proteins~\cite{Balabanian18}. Both in the case of actin filaments~\cite{Tang96,Deshpande12} and in that of microtubules~\cite{Needleman04,Hamon11,Chung16}, bundles of parallel filaments have been shown to form \textit{in vitro} in the absence of cross-linkers under certain experimental conditions. Beyond the cytoskeleton, several enzymes form filaments in cells, with important biological functions, and these filaments also often self-assemble into larger assemblies, especially bundles~\cite{Park19}.
The Casimir interaction considered in this paper matters in particular at dimensions relevant for bundles of actin filaments and of microtubules. Therefore, it should have important implications in
the self-assembly and cohesion of bundles of filaments at the cell scale~\cite{Burla19}. 

Let us note that 
the configuration with metallic cylindrical surfaces 
separated by a vacuum gap 
was proposed as a platform for precision Casimir experiments~\cite{BrownHayes2005,Wei2010,Decca2011,Bsaibes2020}. The Casimir force between crossed cylindrical surfaces in air was measured for distances up to $0.1\,\mu{\rm m}$~\cite{Ederth2000}. 
Theoretical results for metallic
cylindrical surfaces in vacuum were derived at zero~\cite{Emig2006,Bordag2006,Rahi2008,Reid2009,Lombardo2010,Noruzifar2012,Teo2015} and finite temperatures~\cite{Teo2011,Rodriguez-Lopez2012}, as well as in a non-equilibrium configuration~\cite{Golyk2012}.

This work is organized as follows. First, we illustrate by molecular dynamics simulations the fact that transverse modes of the electromagnetic field are not screened by ions, which is crucial to the existence of a long-range Casimir force. Then, we calculate the Casimir interaction between two parallel dielectric cylinders immersed in salted water, using the scattering formalism, and we discuss its universality. Next, we apply our results to bundles of biological filaments, focusing on the specific cases of actin filaments and microtubules. Finally, we discuss the quantitative importance of the electromagnetic Casimir attraction in these biological systems.

\section{Transverse electromagnetic modes are not screened}

\begin{figure}[htbp]
	\includegraphics[width=0.9\columnwidth]{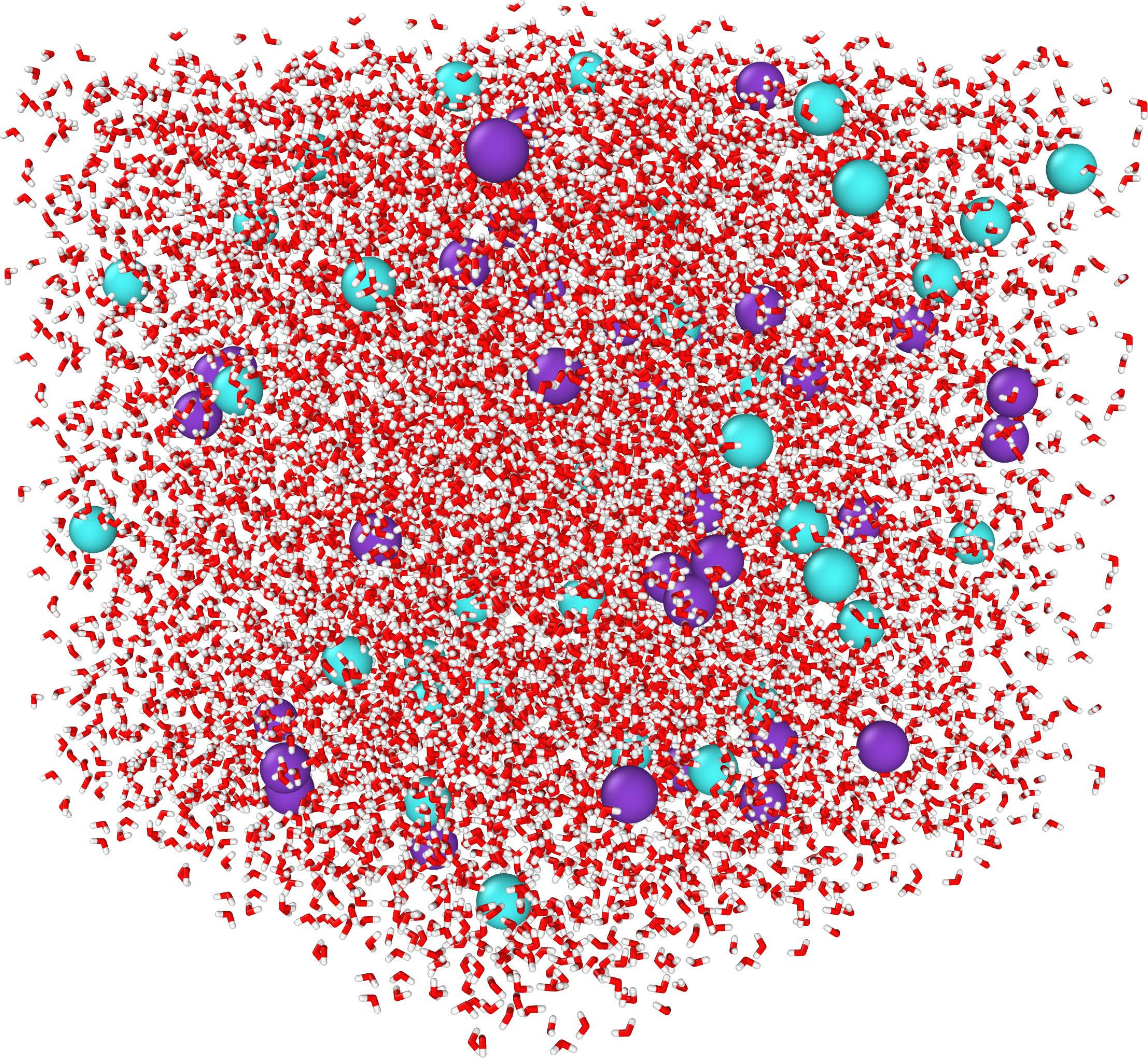}
	\centering
	\caption{\textbf{Snapshot of our molecular dynamics simulation of an electrolyte solution.} The red and white sticks represent water molecules, the light blue spheres represent potassium K$^+$ ions, and the purple spheres bromide Br$^-$ ions. The box is a cube with side length ${\cal L}=6.5\,$nm. Periodic boundary conditions are used. }
	\label{fig:SI2}
\end{figure}

Despite strong screening, the Casimir interaction includes a long-range unscreened part, due to the effect of thermal electrodynamical fluctuations propagating in the medium without being screened~\cite{MaiaNeto2019}. To illustrate this key point, while relying on a molecular description of the environment, we perform molecular dynamics simulations, supported by a classical field theory calculation (see Appendices \ref{appMolecular} and \ref{appClassical}). We simulate pure water using a classical rigid model for water molecules~\cite{azcatl2014}, as well as an electrolyte solution with concentration 0.2 mole per liter of potassium bromide (KBr) (see Fig.~\ref{fig:SI2}). This is in the range of typical cytoplasmic concentrations, and is thus relevant for our applications to bundles of biological filaments below.  

We compute the static dielectric correlation spectrum in Fourier space for these two media. Longitudinal and transverse correlation functions are expressed from the spatial distribution of the charges in the medium and averaged on the simulation time (see Appendix A.2 for details). Figure \ref{fig:SI}(a) shows the longitudinal susceptibility for pure water (blue markers) and for the electrolyte (red markers) for wavevector norms $q\leq1.5\,$\AA$^{-1}$ as we focus on long-range interactions. We observe that the longitudinal susceptibility of the electrolyte significantly differs from the pure water one at low $q$. Figure~\ref{fig:SI}(b) shows the transverse susceptibility for water (blue markers) and for electrolyte (red markers) for $q\leq1.5\,$\AA$^{-1}$. We observe that the transverse susceptibility is not affected by the salt, in agreement with a previous study~\cite{becker2023}.

\begin{figure}[htbp]
	\includegraphics[width=0.95\columnwidth]{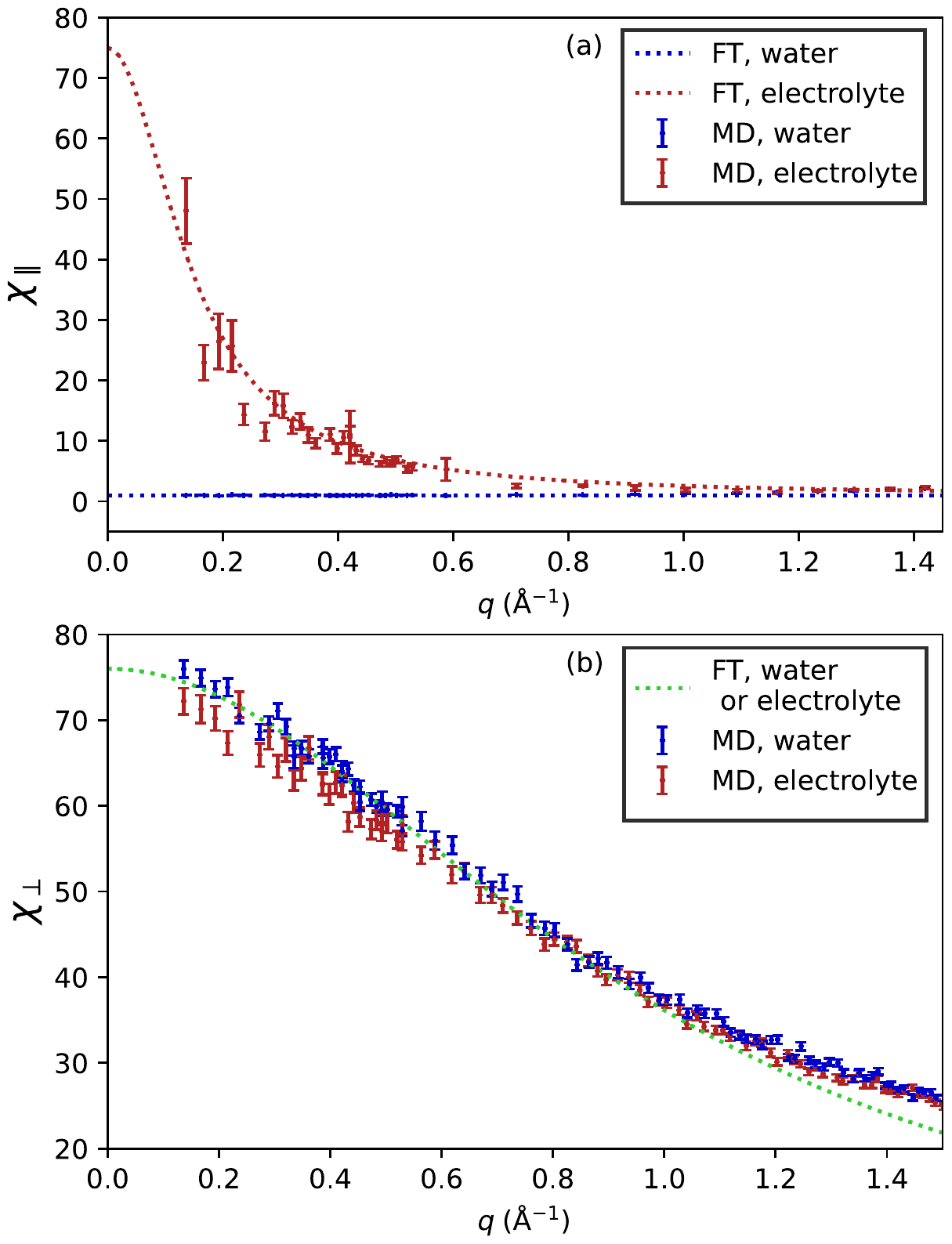}
	\centering
	\caption{\textbf{Susceptibilities for water and electrolyte.} (a) Longitudinal susceptibility $\chi_\parallel(q)$. (b) Transverse susceptibility $\chi_{\perp}(q)$. In both panels, susceptibility is shown as a function of the wavevector norm $q$ (\AA$^{-1}$) for pure water (blue) and for a 0.2 mole per liter KBr electrolyte solution (red). Markers correspond to results obtained from molecular dynamics simulations (MD, see Appendix A for details). Dashed lines correspond to field theory (FT) equations given respectively by Eqs.~(\ref{chiMFmoldyn}) and (\ref{chisalt}) for pure water and for the electrolyte, for parameter values $K=1/76$, $\kappa_t=0.0145\,$\AA$^{2}$ (see Appendix B for derivation). For the transverse susceptibility, the same FT curve is found for both systems (green).  The Debye length associated to the electrolyte solution considered here is $\lambdaD=6.8\,$\AA.}
	\label{fig:SI}
\end{figure}

In addition, we use classical field theory to compute the response functions of electrolytes. We express  the longitudinal and transverse response of pure water using the framework of nonlocal electrostatics~\cite{maggs2006,berthoumieux2015}. We adjust the two parameters of the model to fit the data of molecular dynamics (see Appendix B2 for details). In water, for this range of $q$, the longitudinal  susceptibility is constant (dashed blue line in Fig. \ref{fig:SI}(a)) whereas the transverse one presents a Lorentzian decay (dashed green line in Fig. \ref{fig:SI}(b)). The transverse susceptibility of the electrolyte is unchanged when compared to pure water, as seen in  Fig. \ref{fig:SI}(b). Conversely, the longitudinal one presents a Lorentzian decay induced by Debye screening in the electrolyte, as shown by Eq.~(\ref{chiMFmoldyn}).
A very good agreement is obtained between field theory and simulations, as shown by Fig.~\ref{fig:SI}.  

We checked the robustness of these conclusions by simulating an aqueous electrolyte using another water model and NaCl ions instead of KBr. We indeed obtained similar results, see Appendix \ref{appMolecular}.

We thereby confirm that the longitudinal spectrum is modified by the presence of salt, due to the screening of the correlations in electrolytes beyond the Debye length, whereas the transverse correlation spectrum remains unaffected by the presence of salt. The absence of screening of the transverse modes by salt allows the long-range Casimir force.

\section{Casimir interaction between two dielectric cylinders}

\begin{figure}[htbp]
\centering
\includegraphics[scale=0.3]{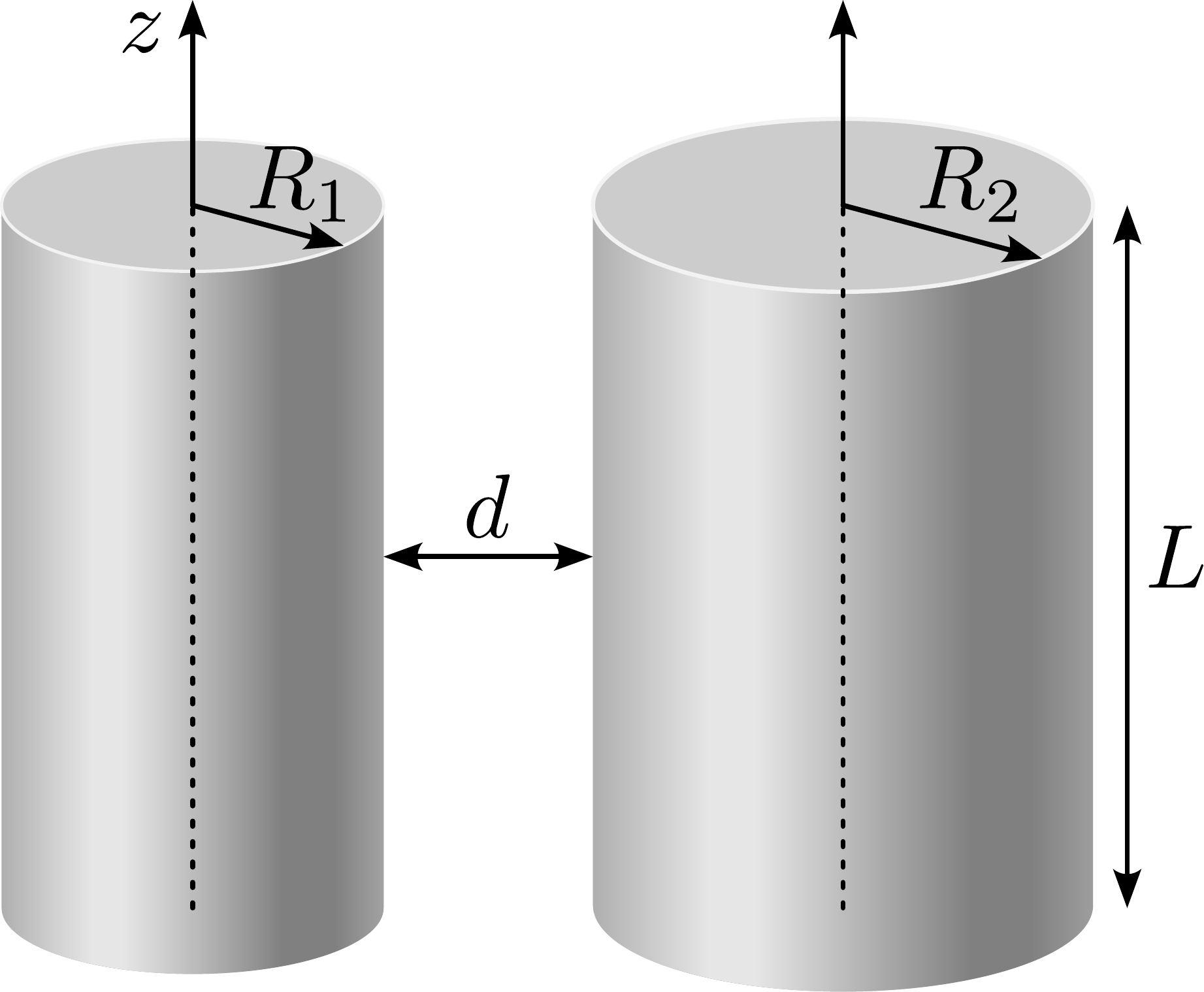}
\caption{\textbf{Geometry studied in this paper.} Two parallel cylinders of length $L$ and radii $R_1$ and $R_2$ are separated by a distance $d$ of closest approach in salted water.}
\label{parallelcylinders}
\end{figure}

We use the scattering formalism \cite{Lambrecht2006,Rahi2009} to calculate the Casimir interaction in salted water at room temperature between two parallel dielectric cylinders with length $L$ and different radii $R_1,R_2$ separated by a distance $d$ of closest approach. 
The cylinder-plane configuration is included in the calculation, for an infinite second radius $R_2$.
We focus on cylinders much longer than the separation distance ($L\gg d$, see Fig.~\ref{parallelcylinders}), thus neglecting edge effects.  
We consider a salt concentration typical of biological media, with the Debye screening length $\lambdaD$ much smaller than the distance $d$. All electrostatic interactions, as well as contributions to the Casimir energy arising from longitudinal modes~\cite{MaiaNeto2019, Nunes2021}, are then efficiently screened. This is a first reason why the resulting interaction  will be independent of many details of the physical configuration. This universal Casimir interaction arises from transverse modes, which are not screened, as discussed in the previous section. 

Another reason for this universality will become clear when describing the scattering formalism \cite{Lambrecht2006,Rahi2009} employed to compute the Casimir interaction for arbitrary values of the geometrical dimensions. 
In general, the interaction is given by a sum over Matsubara frequencies \cite{DLP1961,Schwinger1978}. The first Matsubara term overtakes all other ones when thermal fluctuations dominate~\cite{Parsegian1971}, which is the case for filaments at the cell scale at physiological temperature. This first term corresponds to electromagnetic response functions evaluated at zero frequency. As salted water features an ionic conductivity leading to a divergence of its contribution to the dielectric response, the resulting interaction does not depend on the detailed dielectric function of the cylinders. 

Within the scattering approach, the zero-frequency term giving the thermal Casimir interaction energy is written as an integral over the wavevector $k$ along the $z$-axis of the cylinders (see Fig.~\ref{parallelcylinders})
\begin{equation}
\label{scaform}
\caF = \kB T\,L\, \int_0^{\infty}\frac{\di k}{2\pi}\,\log \det 
\left[\caI-\caM(k)\right]~.
\end{equation}
Here $\caI$ is the unit matrix and $\caM$ the round-trip operator after scattering on the two cylinders
\begin{equation}
\label{M}
{\caM}= {\caR}_1\,{\caT}_{12}\, {\caR}_2\,{\caT}_{21}   ~.
\end{equation}
The latter accounts for reflection operators $\caR_j$ on each cylinder ($j=1,2$), the translation operator $\caT_{21}$ from the axis of cylinder 1 to the axis of cylinder 2, and the reciprocal translation operator $\caT_{12}$ (from the axis of cylinder 2 to the axis of cylinder 1).

The round-trip operator $\caM$ can be written in terms of cylindrical modes associated to given values of $k$ and of an integer number $m$ denoting the angular momentum component along the symmetry axis for each cylinder. 
The zero-frequency term (\ref{scaform}) giving the thermal Casimir interaction is calculated at the static limit for all reflection operators. As a consequence of the Debye screening mechanism, only the transverse-magnetic cylindrical modes contribute (TM modes with the magnetic field perpendicular to the symmetry axis). 

Due to the rotational symmetry of each cylinder, the reflection operators are diagonal in the representation defined by the cylindrical modes. The corresponding matrix elements are evaluated by taking into account the finite conductivity of salted water due to the ions in solution. 
As the dielectric permittivity of salted water diverges in the limit of zero frequency, the matrix elements do not depend on the dielectric response of the cylinder material. 
Using the known reflection matrix for cylinders~\cite{Bohren1998}, we derive for our geometry  ($j=1,2$ for the two cylinders)
\begin{equation}
\label{rTMzero}
\langle m,{\TM}| {\caR}_j|m,{\TM}\rangle =
   -\frac{\imath\pi}{2}\,(-1)^m\,\frac{I_m'(kR_j)}{K_m'(kR_j)}~, 
\end{equation}
where $I_m(x)$ and $K_m(x)$ are the modified Bessel functions of the first and second kinds, respectively (\S 10-25 in~\cite{DLMF}).

The distance between the axes of the two cylinders shown as dotted lines in Fig.~\ref{parallelcylinders} is $D=d+R_1+R_2$. 
Translations along the $x$-axis are described by the following matrix elements 
\begin{equation}
\label{translationzero}
\begin{aligned}
&\langle m',{\TM}| {\caT}_{21}|m,{\TM}\rangle=
   \frac{-2\imath}{\pi}\,(-\imath)^{m-m'}\,K_{m-m'}(kD) ~,\\
&\langle m',{\TM}| {\caT}_{12}|m,{\TM}\rangle=
   \frac{-2\imath}{\pi}\,\imath^{m-m'}\,K_{m-m'}(kD)   ~,
\end{aligned}
\end{equation}
where we used Graf's addition theorem for Bessel functions (\S 10-23 in~\cite{DLMF}).

Explicit results for the Casimir free energy are obtained by combining 
(\ref{scaform}-\ref{translationzero}). We write the result as follows:
\begin{equation}
\label{expli3}
{\caF}=-\kB T \frac{L}{d}\,\phi\left(d,R_1,R_2\right) ~.
\end{equation}
The free energy $\caF$ is thus proportional to the cylinder length $L$, with the latter measured as the dimensionless number $L/d$. The dimensionless quantity $\phi$ only depends on the radial dimensions $d,R_1,R_2$ characterizing the two-cylinder system. In fact, $\phi$ only depends on the two ratios of these three dimensions. Thus, $\caF$ does not depend on any material properties of the cylinders or of the surrounding fluid, and is universal. A convenient representation is 
\begin{equation}
\label{defx}
\begin{aligned}
&\phi \equiv \phi_u\left(x\right)~,
\textrm{ with } u=\frac{R_1R_2}{(R_1+R_2)^2}~, \\
&\quad x=\frac{d}{\Reff}~,\quad \Reff=\frac{R_1\,R_2}{R_1+R_2}~.
\end{aligned}
\end{equation}
The parameter $u$ is a symmetrized ratio of the two radii. It runs from $u=0$ in the cylinder-plane geometry to $u=1/4$ for equal radii. 
Meanwhile, $x$ compares the distance $d$ of closest approach to the effective radius $\Reff$. The latter is equal to the cylinder radius in the cylinder-plane geometry, and to $R/2$ for cylinders with equal radii $R$. 
Note that the case of cylinders of equal radii corresponds to the quantity $\phi_u$ in Eq.~\eqref{defx} calculated for $u=1/4$, and written as a function of $d/R\equiv x/2$.

\begin{figure}[htbp]
\centering
\includegraphics[scale=0.82]{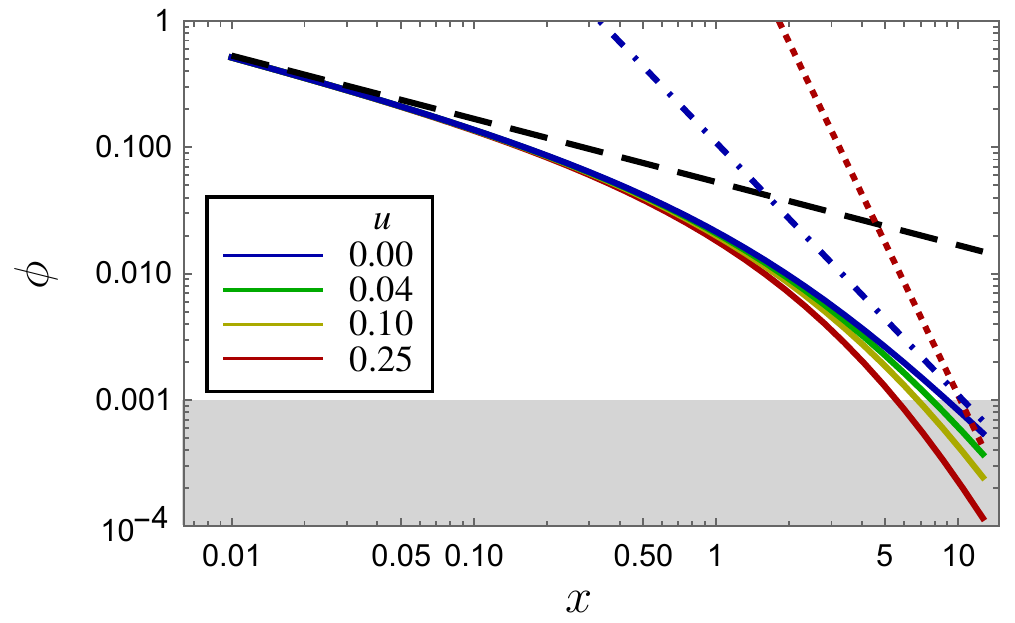}
\caption{\textbf{Universal function $\phi_u$ versus $x=d/\Reff$.} The parameter $u$ takes values $0,0.04,0.1, 0.25$ respectively from the top curve to the bottom one. The first and last cases correspond to cylinder-plane and equal-radii cylinders.
The full results (solid curves) agree with the proximity force approximation (PFA) expression (dashed line) at short separations and with approximations discussed in the text (dotted and dash-dotted lines) at long distances. In the grayed out band, the value of $|\caF|$  (calculated with $L/d=10^3$) is dominated by the Brownian motion energy scale $\kB T$. }
\label{numerical_results}
\end{figure}

In Fig.~\ref{numerical_results}, we plot the function $\phi_u(x)$ for four different values of $u$.
The Casimir interaction energy per unit length $\caF/L$ is obtained by multiplying  $\phi_u$ by $-\kB T/d$ (note that $\caF<0$ and $\phi_u>0$). 
We optimized the numerical evaluation by expanding the round-trip operator in the plane wave basis rather than in the cylindrical one \cite{Spreng2018, Schoger2021}.  
Explicit expressions for the scattering matrix elements in the plane wave basis are readily derived from the results presented above. 
Since the wave-vector is a continuous variable, the determinant in (\ref{scaform}) is calculated with the help of Nystr\"om discretization, as in the calculation of the Casimir interaction between spheres~\cite{Spreng2020}.
 To facilitate applications, we provide the numerical evaluations of $\phi$ on a repository (see below, section ``Code and data availability") for the four values of $u$ shown on Fig.~\ref{numerical_results}, over the domain $0.1<x<15$, which should be appropriate for most applications.
 
Our full numerical results hold for arbitrary values of distance. Let us compare them with the proximity force approximation (PFA) or Derjaguin approximation~\cite{Derjaguin1934} in the limit $x \ll 1$. This approximation amounts to replacing the function $\phi_u$ by the result for parallel planes averaged over the local distances between the cylindrical surfaces, yielding
\begin{equation}
\label{PFAres}
\phi_{\rm PFA}(x) = \frac{H}{24}\,\sqrt{\frac2x}~, \textrm{ with } ~H=\frac34 \,\zeta(3)\,\kB T ~.
\end{equation}
The effective Hamaker coefficient ${\displaystyle H\approx 0.9\,\kB T}$ is the one calculated for dielectric surfaces separated by salted water~\cite{MaiaNeto2019}, with $\zeta(3)\approx 1.202$ denoting Ap\'ery's constant.
The PFA is indicated as a dashed line in Fig.~\ref{numerical_results}, and is indeed a good approximation of numerical results at short distances, whereas it increasingly overestimates the magnitude of the interaction energy as $x$ increases. Note that the PFA expression \eqref{PFAres} does not depend on $u$.

Analytical results can also be derived in the opposite limit $x\gg1$. Indeed, the single round-trip approximation ${\displaystyle \log \det \left({\caI}-{\caM}\right)\approx -{\rm Tr}\,{\caM}}$ is then sufficient to get an estimate of the free energy (\ref{scaform}). We thus find two different results for the case of two cylinders ($u>0$) and for the cylinder-plane geometry ($u=0$)
\begin{equation}
\label{long_distance}
\begin{aligned}
&\phi_u(x)\approx \frac{891\pi}{4096\,u^2\,x^4}\,,\quad x\gg1\,,\quad u>0~, \\
&\phi_0(x)\approx \frac{7}{64\,x^2}\,,\quad x\gg1\,,\quad u=0~. 
\end{aligned}
\end{equation}
The comparison between these two long-distance results indicates that the reduction with respect to the PFA limit is stronger for two cylinders than for a cylinder and a plane, as expected. 
Formulas in the first and second lines of Eq.~\eqref{long_distance} 
are shown respectively as dotted (for $u=1/4$) and dash-dotted lines in Fig.~\ref{numerical_results}.

A key feature of the results obtained here is their universality. Our results are valid for whatever dielectric response functions of the cylinders. They depend only on the dimensionless length-to-distance ratio $L/d$ and on the dimensionless ratios of the radial dimensions $d,R_1,R_2$. Furthermore, we show in Appendix \ref{appConformal} that most of the dependence on radial dimensions is captured by considering the dimensionless free energy $\phi$ as a function of a conformally invariant geometrical parameter.

\section{Application to bundles of biological filaments}

How relevant is the universal Casimir attraction between dielectric cylinders in biological systems? To address this question, it is important to compare the magnitude of the Casimir free energy to the thermal energy scale $\kB T$. For actin filament bundles and for microtubule bundles, the filament length $L$ will be in the micrometer range while the inter-filament distance $d$ will be in the nanometer range (see below). Thus, typically, the ratio $L/d$ of cylinder length over separation distance is of the order of $10^3$ for bundles in cells. With such a value of $L/d$, finite-size (edge) effects are expected to be negligible, which is consistent with our assumptions. The grayed out band on the lower part of Figure~\ref{numerical_results} shows the domain where the Casimir free energy $\caF$ is overtaken by $\kB T$ for $L/d\sim 10^3$. Importantly, we find that the  Casimir binding energy is larger or of the same order as the thermal scale in a broad range, namely $x<5$, as indicated by Fig.~\ref{numerical_results}. As expected from the fact that the Casimir energy is proportional to the length of the cylinders, the range where the Casimir force plays an important role is much broader in the two-cylinder geometry ($x<5$) than in the two-sphere geometry (where it was evaluated as $x<0.1$ at the end of \cite{Schoger2022}). 

The values of $x$ corresponding to the case of the filament bundles in cells discussed below are such that the Casimir force should play an important role in these systems. In addition, these practically relevant values of $x$ also lie right in the crossover between the PFA and the long-distance limits shown in Fig.~\ref{numerical_results}. In this intermediate range, both short- and long-distance approximations overestimate the exact energy that we  computed numerically by approximately one order of magnitude. This highlights the importance of our calculation and of our full numerical results for these applications. The results obtained here are therefore of importance for the self-assembly and cohesion of filament bundles in cells, with implications for cellular and molecular biology.  

\begin{figure}[htbp]
    \centering
    \includegraphics[width=0.5\textwidth]{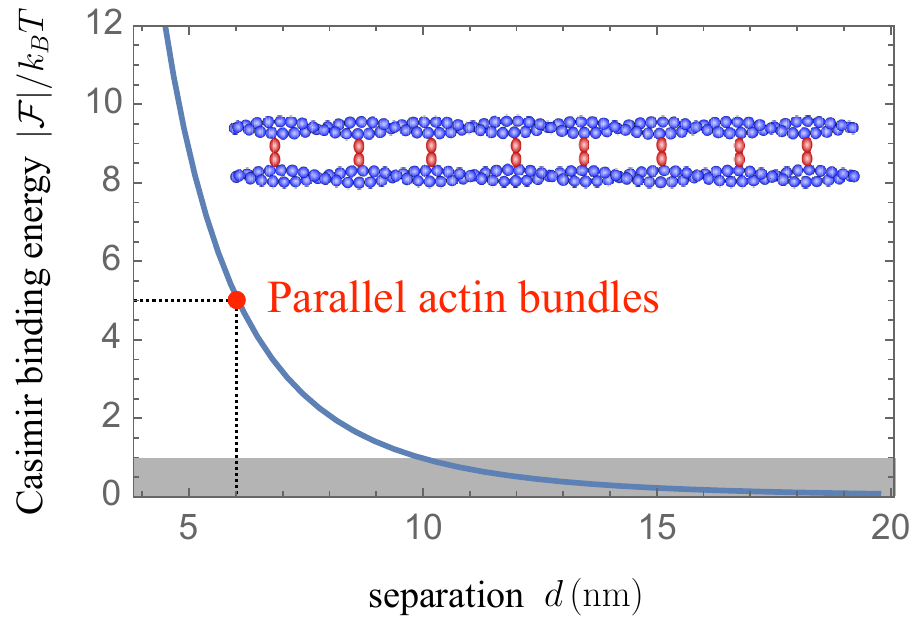}
    \caption{ \textbf{Casimir binding free energy between actin filaments.}
   The Casimir binding free energy $|{\cal F}|=-{\cal F}$ (from Eq.~\ref{expli3}) is shown versus the separation distance $d$~(nm) for two parallel actin filaments. Each filament has a radius $3\,{\rm nm}$ and length $15\,\mu{\rm m}$. $|{\cal F}|$ is expressed in units of $\kB T$, and the zone where $|{\cal F}|<\kB T$ is shaded in gray. Casimir binding free energies above the gray zone are expected to matter in practice, including in the physiological case of parallel actin bundles (red marker). Inset: schematic of two actin filaments (blue) in a parallel actin bundle. Cross-linkers are shown in red.}
    \label{fig:fig2}
\end{figure}

Let us now assess more precisely the magnitude of this interaction in the specific case of actin bundles. Actin filaments are double helices of homopolymers of monomeric actin. They can be approximately described as cylinders with a radius $R$ around $3\,$nm. They form bundles in cells, where actin filaments are cross-linked by specific proteins into arrays of parallel filaments. In parallel actin bundles, which support projections of the cell membrane such as microvilli, microspikes or filopodia, actin filaments (assembled with fimbrin, fascin or villin) are approximately $d=6\,$nm apart (note that we use the closest approach distance $d$ here and throughout)~\cite{Claessens08,Volkmann01}. In this case, Eq.~\ref{expli3} yields a Casimir binding free energy per unit length of $|{\cal F}|/L = 0.33\, \kB T /\mu{\rm m}$, giving the substantial value $|{\cal F}|=5\, \kB T$ for a length $L= 15\,\mu{\rm m}$, which is on the order of the size of a cell and of the persistence length of actin filaments~\cite{Brangwynne07}. Such a value, significantly larger than the scale $\kB T$ of thermal fluctuations, demonstrates the practical relevance of the Casimir interaction between actin filaments in the physiological configuration of parallel bundles. In contractile bundles, which are present in stress fibers, and in the mitotic contractile ring, actin filaments (assembled with alpha-actinin) are separated by $d=33\,$nm~\cite{Volkmann01}, yielding a smaller value of $|{\cal F}|\sim 10^{-2}\, \kB T$ for a length $L= 15\,\mu{\rm m}$, which is not relevant as it is well below the scale $\kB T$ of thermal fluctuations. In Fig.~\ref{fig:fig2}, we show the Casimir binding free energy $|{\cal F}|=-{\cal F}$ versus the separation $d$ for actin filaments with $L= 15\,\mu{\rm m}$. The Casimir interaction then exceeds the scale of thermal fluctuations for separations $d\lesssim 10\,$nm, which includes parallel bundles but not contractile bundles.

\begin{figure}[htbp]
    \centering
    \includegraphics[width=0.5\textwidth]{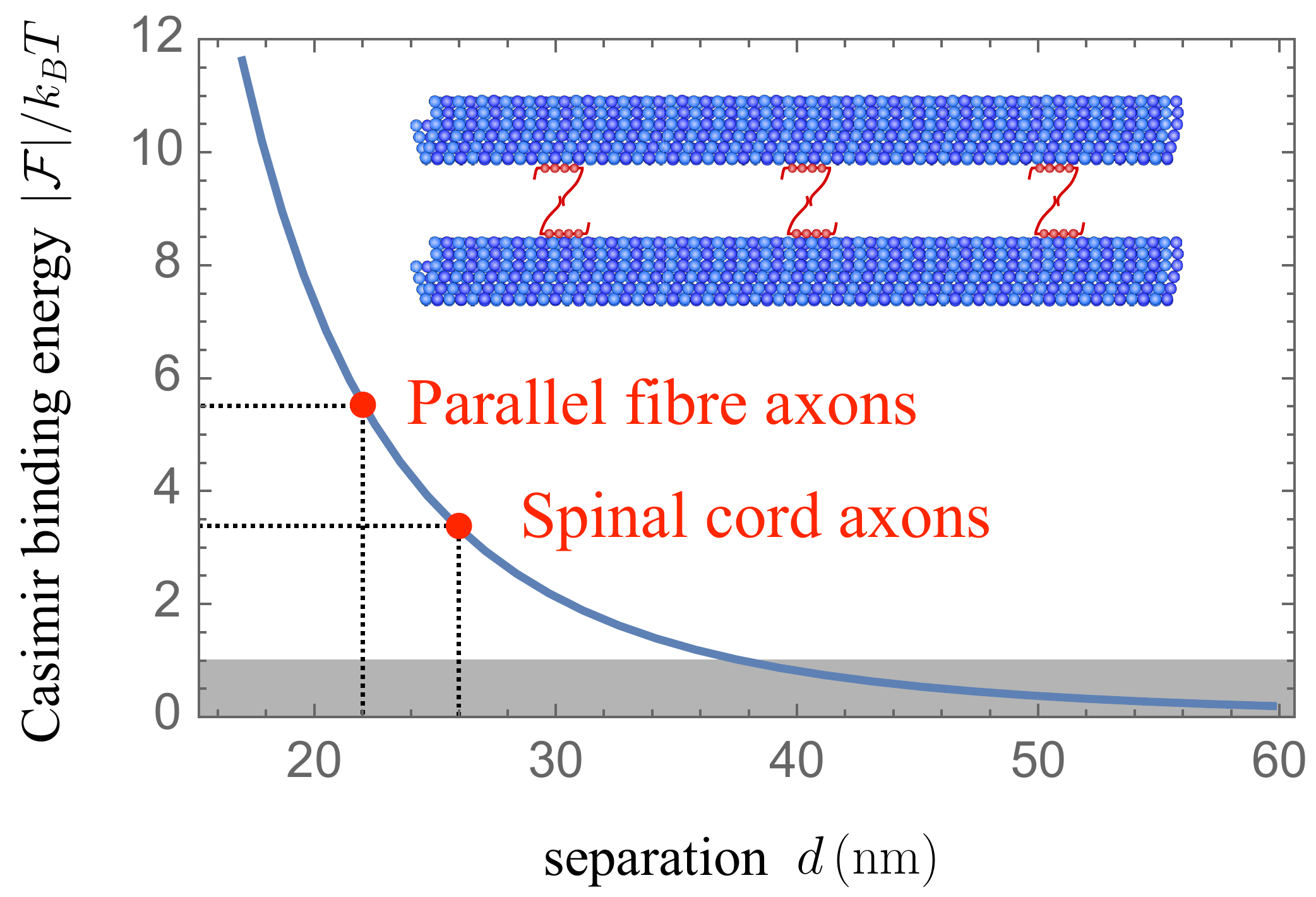}
    \caption{ \textbf{Casimir binding free energy between microtubules.}
    Same as in Fig.~\ref{fig:fig2}, but in the case of two parallel microtubules, each with radius $12\,{\rm nm}$ and length $50\,\mu{\rm m}$. The physiological cases of parallel fibre axons and spinal cord axons are shown by red markers. Inset: schematic of two microtubules (blue) in a bundle. Cross-linkers are shown in red.}
    \label{fig:fig3}
\end{figure}

Another biologically important system where Casimir interactions between filaments are relevant regards microtubule bundles. Microtubules can be viewed as cylinders with a radius of about $12\,$nm. They can grow as long as $50\,\mu$m and their persistence length is around $1\,$mm~\cite{Hawkins10}. They form bundles where the separation between neighboring microtubules is set by microtubule-associated proteins, of which various types exist~\cite{Balabanian18}. 
Plant cells often possess large arrays of parallel microtubules, whose alignment is maintained over the whole cell, and where separations are similar to the microtubule diameter~\cite{Chan99,Gaillard08}. Microtubule bundles are also present in neurons, where they play important roles, and the separations between adjacent microtubules in Purkinje cell dendrites, parallel fibre axons and white matter spinal cord axons were found to be $64\pm 10\,$nm, $22\pm 10\,$nm and $26\pm 10\,$nm, respectively~\cite{Chen92}.
For $d=22\,{\rm nm}$, we find $|{\cal F}|/L \approx 0.112\, \kB T /\mu{\rm m}$, giving $|{\cal F}|=5.6\, \kB T$ for $L= 50\,\mu{\rm m}$ and $|{\cal F}|=1.7\, \kB T$ for $L= 15\,\mu{\rm m}$. In Fig.~\ref{fig:fig3}, we show the Casimir binding free energy $|{\cal F}|$ versus the separation $d$ for microtubules with $L= 50\,\mu{\rm m}$. The Casimir interaction then exceeds the scale of thermal fluctuations for separations $d\lesssim 35\,$nm, which includes the physiological separations found in parallel fibre axons and white matter spinal cord axons, as well as in plant cells, but not in Purkinje cell dendrites.

\section{Discussion}

Actin bundles and microtubule bundles are typically held in place by cross-linking proteins in cells. However, in electrolyte solutions containing polycations e.g. Mg$^{2+}$, actin filaments can form bundles \textit{in  vitro} in the absence of cross-linking proteins~\cite{Tang96,Deshpande12}. Microtubule bundles can also self-assemble \textit{in vitro} above a certain concentration of multivalent cations~\cite{Needleman04,Hamon11}. This demonstrates that the tendency of these filaments to self-assemble is quite generic. In this light, cross-linkers could help maintain spacing between filaments~\cite{Mephon16}. Interestingly, beyond cytoskeletal proteins, multiple enzymes form filamentous structures within cells, which then assemble into large-scale self-assembled structures (foci, rods, rings, sometimes called cytoophidia), which are membraneless and reversible. Enzyme filamentation is associated to multiple functions in cells, including determining cell shape or regulating enzyme activity~\cite{Park19}. Enzymes that form filaments and higher-order structures in cells include acetyl CoA carboxylase (ACC), CTP synthetase (CtpS)~\cite{Ingerson10,Barry14}, inositol monophosphate dehydrogenase (IMPDH)~\cite{Juda14,Johnson20}, and many others~\cite{Park19}. Beyond enzymes, it was recently shown that mutating proteins that spontaneously form symmetric homo-oligomers can lead to polymerization in a quite generic manner. Furthermore, these mutant proteins very often form larger structures such as fibers or foci, and some were shown to bundle~\cite{Garcia17,Garcia22}. These findings hint at a general trend of filaments to self-assemble into higher-order structures in cells.

While the universal attractive interaction discussed here should play a key role in these various bundles, these systems are complex and involve many other interactions. Electrostatic interactions are screened at the usual separations involved in actin bundles and microtubule bundles, but they matter at shorter separations. For instance, the surface of filamentous actin is overall negative, with a highly heterogeneous charge distribution, leading to subtle collective dynamics of counterions close to actin filaments~\cite{Angelini06}. In addition, the depletion interaction~\cite{Asakura54,Asakura58,Vrij76,Marenduzzo06}, which arises from excluded volume effects on crowding agents, is important in cells due to how crowded the cytoplasm is. Indeed, around 30\% of its volume is estimated to be occupied by macromolecules~\cite{Zimmerman91,Ellis01,Marenduzzo06}, with notable heterogeneities~\cite{Gnutt15,Rivas16}. The range of the depletion interaction is given by the diameter of a typical depletant, which corresponds to macromolecules such as globular proteins in cells, with diameters of order $5\,$nm~\cite{Marenduzzo06}. Depletion interactions have been studied experimentally in controlled in-vitro systems where the concentration of depletant polymers can be tuned. For instance, an adhesion strength of $7\,k_\mathrm{B}T/\mu$m was measured between sickle hemoglobin fibers in a solution of monomeric hemoglobin~\cite{Jones03}, while the attractive interaction between two actin filaments in a solution of depletant polymers was found to be of order of a few times $10\,k_\mathrm{B}T/\mu$m~\cite{Lau09}, and a similar value was found between two microtubules~\cite{Hilitski15}. Thus, this interaction is strong between parallel filaments in a cell, but it is also very short-ranged, with a range of order $5\,$nm.

What sets the Casimir interaction we calculated apart from electrostatic and depletion interactions, and to our knowledge, from all other interactions at equilibrium, is its long range, which arises from the lack of screening of transverse electromagnetic fluctuations. An out-of-equilibrium long-range fluctuation-induced interaction was recently predicted between neutral objects immersed in  electrolytes subject to an external electric field \cite{Golestanian2021}. This force can be of importance at the cell scale, e.g. for ion channels. More generally, Casimir forces present interesting out-of-equilibrium properties~\cite{Dean12, Dean16}. Such effects could be all the more important for cytoskeletal filaments that the cytoskeleton is an active system~\cite{Julicher07,Gladrow16,Gross19}. Here, we showed that an equilibrium long-range universal interaction exists between filaments in cells. 

The Casimir interaction is highly dependent on the geometry of the interacting objects because it arises from the perturbation of electromagnetic fluctuations by the interacting objects. Here, we showed that its magnitude is several times the scale of thermal fluctuations in the geometry of two parallel filaments whose length (micrometer-scale) is much larger than their radius and separation (nanometer-scale). This is stronger than between two spheres~\cite{Schoger2022}, because long cylinders have a stronger confining effects on electromagnetic fluctuations than spheres. Accordingly, within a cell, the Casimir interaction can be strong between long semi-rigid biopolymers such as those considered here, but is weak between globular proteins. Note that similar geometric effects exist for other fluctuation-induced interactions, e.g. in the case of Casimir-like interactions induced by the thermal fluctuations of the shape of a biological membrane: these interactions are stronger between long parallel rods adsorbed on a membrane~\cite{Golestanian96,Bitbol11} than between circular or point-like inclusions modeling transmembrane proteins~\cite{Goulian93,Fournier97}. Critical Casimir forces can also be important for cylindrical particles immersed in critical binary mixtures~\cite{Trondle10,Labbe14,Labbe17}. In addition to parallel cylinders, another biologically relevant case where the electromagnetic Casimir interaction should matter regards stacks of lipid membranes. Modeling them by parallel dielectric planes immersed in salted water, the Casimir binding free energy between two lipid membranes is $|\mathcal{F}|=2.4\times 10^{-2}\kB T\,A/d^2$, where $A$ is the area of the planes and $d$ their separation \cite{MaiaNeto2019}. It should thus oppose the repulsive Helfrich and hydration interactions~\cite{Helfrich78,Cevc1987,leikin1993,Petrache06,Freund12,Wennerstrom14,Lu15}. 

\section{Summary and conclusion}

The long-range part of the Casimir attraction has a universal form between two dielectric objects immersed in salted water. Here, we calculated the Casimir interaction in the case of two long parallel dielectric cylinders, using the scattering formalism. We demonstrated that this interaction takes values substantially larger than the scale of thermal fluctuations, in the important biological cases of actin bundles and microtubule bundles. 

The long range of the Casimir interaction we calculated arises from the lack of screening of transverse electromagnetic fluctuations, which we confirmed by molecular dynamics simulations. It is this long range that makes the Casimir interaction quantitatively important e.g. between actin filaments at the physiological separation found in parallel bundles. It also sets it apart from other equilibrium interactions present in these structures. The Casimir interaction should thus play an important part in the self-assembly of filament bundles in cells.

\section*{Code and data availability}
Code for our numerical calculations of the Casimir force is freely available in the GitHub repository \url{https://github.com/sprengjamin/CasCy}.
Data corresponding to our figures is freely available in the Zenodo repository \url{https://zenodo.org/doi/10.5281/zenodo.7634525}.

\section*{Acknowledgments}

We are grateful to Romain Guérout, Gert-Ludwig Ingold and Tanja Schoger for fruitful discussions. 
A.-F.~B. thanks the European Research Council (ERC) for funding under the European Union’s Horizon 2020 research and innovation programme (grant agreement No.~851173, to A.-F.~B.). 
H.~B. acknowledges funding from the Humboldt Research Fellowship Program for Experienced Researchers and thanks Roland Netz and the Freie Universit{\"a}t, Berlin, for hospitality.
P.~A.~M.~N. thanks Sorbonne Université for hospitality and acknowledges funding from the Brazilian agencies Conselho Nacional de Desenvolvimento Cient\'{\i}fico e Tecnol\'ogico (CNPq--Brazil), Coordenaç\~ao de Aperfeiçamento de Pessoal de N\'{\i}vel Superior (CAPES--Brazil),  Instituto Nacional de Ci\^encia e Tecnologia Fluidos Complexos  (INCT-FCx), and the Research Foundations of the States of Rio de Janeiro (FAPERJ) and S\~ao Paulo (FAPESP).

\appendix

\section{Molecular dynamics computation of the longitudinal and transverse dielectric susceptibilities for pure water and electrolytes}
\label{appMolecular}

\subsection{Molecular dynamics simulation methods}
We consider a cubic water box of side length ${\cal L}=6.5\,$nm composed of $N_w=8967$ water molecules. We simulate both pure water and a 0.2 moles per liter aqueous solution of KBr. The electrolyte solution contains 33 ion pairs in the box. Figure~\ref{fig:SI2} shows the simulation box.
Our focus is on dielectric properties. Thus, rather than simulating a complex solution matching the cytosol composition, we consider a simple solution whose permittivity is well tabulated~\cite{azcatl2014,loche2021} and matches the cytosolic one. Its Debye length, $\lambdaD=0.68\,$nm, is in the typical range of biologically relevant solutions.

Simulations are performed using the \texttt{GROMACS 2021} molecular dynamics simulation package~\cite{abraham2015}. We employ ion force field parameters optimized for ion solvation and ion-ion interaction~\cite{loche2021}. We use Lorentz–Berthelot mixing rules with respect to solvent Lennard-Jones parameters. The integration time step is set to $\Delta t=2\,$fs. We further checked that a time step of $\Delta t=1\,$fs yields similar results for this system, thus confirming that our time step is small enough. Periodic boundary conditions are used in all directions. Long range electrostatics are handled using the smooth particle mesh Ewald (SPME) technique. Lennard-Jones interactions are cut off at a distance $r_{\rm cut}=0.9\,$nm.  A potential shift is used at the cut-off distance. All systems are coupled to a heat bath at $300\,$K using a v-rescale thermostat with a time constant of $0.5\,$ps. We use the Python librairie \texttt{MDAnanlysis} to treat the trajectories. After creating the simulation box, we perform a first energy minimization. Specifically, we equilibrate the system in the NVT ensemble for $200\,$ps, and afterwards in the NPT ensemble for another $200\,$ps using a Berendsen barostat at $1\,$bar.
Production runs are then performed in the NVT ensemble for $20\,$ns. 

We performed simulations with the TIP4P/$\epsilon$ water model~\cite{azcatl2014}, a 4 interaction site,  three point-charges and one Lennard Jones reference site model. The Lennard-Jones (LJ) center is placed on the oxygen atom. Charges are placed on the hydrogen atoms and on an additional interaction site, M, carrying the negative charge. The ions (K$^+$ and Br$^-$) were treated according to the force field developed in~\cite{loche2021}.

\begin{figure}[htbp]
	\includegraphics[width=\columnwidth]{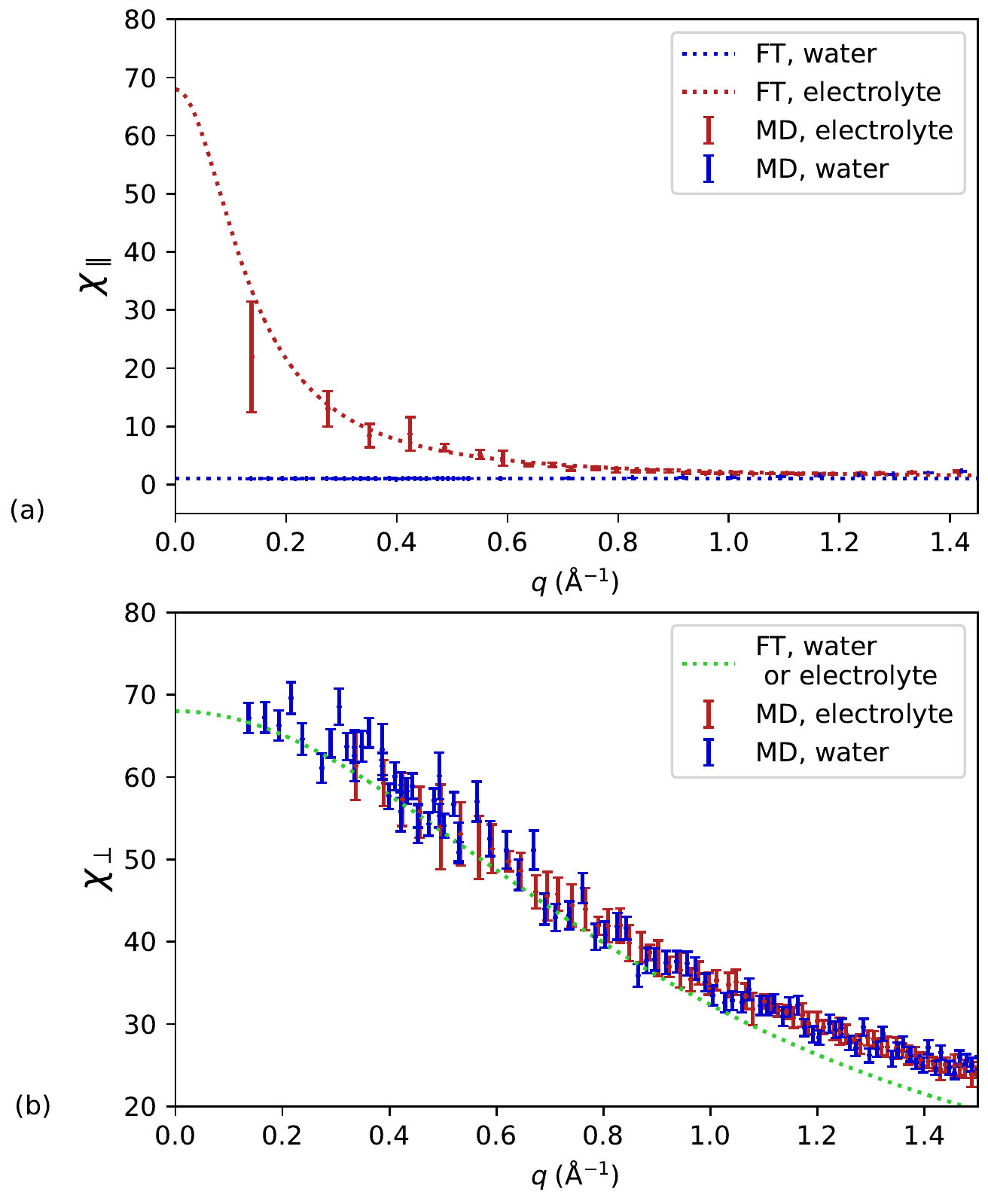}
	\centering
	\caption{\textbf{Susceptibilities for SPC/E water and NaCl electrolyte.} (a) Longitudinal susceptibility $\chi_\parallel(q)$. (b) Transverse susceptibility $\chi_{\perp}(q)$. In both panels, susceptibility is shown as a function of the wavevector norm $q$ (\AA$^{-1}$) for pure water (blue) and for a 0.15 mole per liter NaCl electrolyte solution (red). Markers correspond to results obtained from MD simulations. Dashed lines correspond to field theory (FT) equations given respectively by Eqs.~(\ref{chiMFmoldyn}) and (\ref{chisalt}) for pure water and for the electrolyte, for parameter values $K=1/68$, $\kappa_t=0.0145\,$\AA$^{2}$ (see Appendix B for derivation). These parameters were adjusted to reproduce the dielectric properties of SPC/E water. For the transverse susceptibility, the same FT curve is found for both systems (green).  The Debye length associated to the electrolyte solution considered here is $\lambdaD=7.5\,$\AA.}
	\label{fig:spce}
\end{figure}

As a check, we also performed a simulation with the broadly used SPC/E water model of a 0.15 moles per liter aqueous solution of NaCl, using the force fields in~\cite{loche2021}. We followed the protocol described above. The corresponding results are presented in Fig.~\ref{fig:spce}. Panel (a) confirms the screening of longitudinal modes by the salt, and panel (b) shows that transverse modes remain unaffected. This is fully consistent with our results in Fig.~\ref{fig:SI}. Therefore, our results are robust to changing the electrolyte and the water model.

\subsection{Computation of the susceptibilities}
To compute the dielectric susceptibility $\chi(\vec{q})$ in Fourier space, we use the fluctuation-dissipation theorem, relating $\chi(\vec{q})$ to the fluctuations of the polarization field $\bm{\mathcal{P}}$, as follows,
\begin{equation}
	\label{FDT}
\langle \bm{\mathcal{P}}(\vec{q}) \bm{\mathcal{P}}(\vec{\vec{-q}}) \rangle=\epsilon_0 \kB T \chi(\vec{q})\,.
\end{equation}
Using the isotropy and the homogeneity of the medium, the susceptibility can be decomposed in a longitudinal part $\chi_\parallel(q)$ and a transverse part $\chi_\perp(q)$ as follows:
\begin{eqnarray}
&&\chi_{ij}(q)=\chi_{\parallel}(q)\frac{q_iq_j} {q^2}+\chi_\perp(q)\left(\delta_{ij}-\frac{q_iq_j}{q^2}\right) \notag\\
&&\quad\textrm{where } (i,j)\in\{x,y,z\}^2\,.
\label{chiMFmoldyn}
\end{eqnarray}

\paragraph*{Longitudinal susceptibility.} The local partial charge $\rho$ of water obeys  $\rho(\vec{r})=-\bm{\nabla} \cdot \bm{\mathcal{P}}(\vec{r})$. Using this relation, one can express the longitudinal susceptibility as a function of the charge structure factor $S(q)$:
\begin{equation}
	\chi_\parallel(q)=\frac{S(q)}{q^2 \epsilon_0 \kB T}\,.\label{chiparallel}
\end{equation}
The charge structure factor in Fourier space can be decomposed into an intramolecular and an intermolecular part,
\begin{equation}
	S(q)=S_{\rm intra}(q)+S_\mathrm{inter}(q)\,.
\end{equation}
The intermolecular contribution reads
\begin{equation}
	S_{\rm inter}(q)=\frac{4 n_w z^2 e^2}{q^2}\left[h_{\rm MM}(q)+h_{\rm HH}(q)-2h_{\rm HM}(q)\right]\,,
\end{equation}
where $z$ denotes valency, $e$ the elementary charge, $n_w$ the molecular number density, while $h_{\rm IJ}$ is the Fourier transform of $g_{\rm IJ}(r)-1$, $g_{\rm IJ}(r)$ being the radial distribution function associated with the atom couple IJ. 
Next, the intramolecular contribution can be written as
\begin{equation}
	S_{\rm intra}(q)=\frac{4 n_w z^2 e^2}{q^2}\left(\frac{\sin(q d_{\rm HH})}{q d_{\rm HH}}-4\frac{\sin(q d_{\rm HM})}{q d_{\rm HM}}+3\right)
\end{equation}
where $ d_{\rm IJ}$ is the intramolecular distance between atoms I and J. 
At low $q$, the accuracy of this expression of the structure factor decreases, because the function $h_{\rm IJ}(r)$ is obtained at a finite range, imposed by the box size.  To solve this problem, we proceed as follows. 
For $q< 2.5\,$\AA$^{-1}$, we take into account the periodicity of the system by calculating the charge structure factor for discretized values of the wavevector norm $q$, namely $q=2\pi\sqrt{n_x^2+n_y^2+n_z^2}/{\cal L}$, where $(n_x,n_y,n_z)$ are non-negative integers. We then compute directly the charge structure factor from the charge distribution $	\tilde{\rho}(q)$ in Fourier space and its correlations~\cite{kornyshev1996}.

\paragraph*{Transverse susceptibility.} The transverse susceptibility is computed following~\cite{kornyshevtrans}.
The polarization of the medium in Fourier space, namely
\begin{equation}
	{\bf P}({\bf q})=\sum_j {\bf p}_j({\bf q})e^{-\imath {\bf q}\cdot  {\bf r}_j}\,,
\end{equation}
can be written as a sum over the molecular polarization ${\bf p}_j({\bf q})$ of molecule $j$, which reads
\begin{equation}
	{\bf p}_j({\bf q})=	\frac{1}{\sqrt{V}}\sum_\alpha \frac{e\, z_\alpha\, {\bf \delta r}_{\alpha j}}{\imath\, {\bf q}\cdot {\bf \delta r}_{\alpha j } }\left(1-e^{-\imath\,{\bf q}\cdot {\bf \delta r}_{\alpha j }}\right)\,,
\end{equation}
where ${\bf \delta r}_{\alpha j } $ denotes the distance between the charge $\alpha$ and the center of mass of the molecule. 
We then take the transverse part of the polarization ${\bf P}_\perp({\bf q})={\bf q}\times {\bf P}(\bf q)/q$, and compute the transverse susceptibility as
\begin{equation}
	\label{chiperpMD}
	\chi_\perp(q)=\frac{\langle{\bf P}_\perp({\bf q})\cdot{\bf P}_\perp({\bf -q})\rangle}{2 \kB T\epsilon_0}.
\end{equation}
Note that we replace $\left(1-e^{-\imath{\bf q}\cdot {\bf \delta r}_{\alpha j }}\right)/(\imath\,{\bf q}\cdot {\bf \delta r}_{\alpha j }) $ by 1 if $ {\bf q}\cdot {\bf \delta r}_{\alpha j } <10^{-5}$ to prevent numerical errors.\par 

For the longitudinal and transverse susceptibilities, the error bars shown in Fig.~\ref{fig:SI} were derived following the reblocking method \cite{flyvbjerg1989}.

\vspace{0.5cm}

\section{Classical field theory interpretation}
\label{appClassical}

To better understand why longitudinal and transverse fluctuations are  differently affected by salt, we use a classical field theory model for water and electrolytes.

\subsection{Water as a nonlocal dielectric medium}
We describe water as a continuous nonlocal and linear dielectric medium~\cite{maggs2006}. The electrostatic energy $\mathcal{U}_{\rm el}$ of the medium is written as a functional of the polarization field ${\bm{\mathcal P}}$ as follows:
\begin{eqnarray}
\label{HP}
 &&\mathcal{U}_{\rm el}[\bm{\mathcal{P}}]=  \frac{1}{2}\int d{\vec r} d{\vec r} '\frac{\bm{\nabla}_{\vec{r}} \cdot \bm{\mathcal{P}}(\vec{r})\bm{\nabla}_{\vec{r}'}   \cdot \bm{\mathcal{P}}(\vec{r}')}{4 \pi \epsilon_0|\vec{r}-\vec{r}'|} \notag\\
&& +\frac{1}{2 \epsilon_0}\int d\vec{r}\left[K \bm{\mathcal{P}}(\vec{r})^2+\kappa_t(\bm{\nabla} \times  \bm{\mathcal{P}}(\vec{r}))^2\right] \,.
\end{eqnarray}
where the second term has been be expanded following a Landau-Ginzburg approach to encode the correlations of the fluid at the nanoscale. $K$ is a parameter defining the bulk ($q=0$) properties of the medium, and $\kappa_t$ a Landau-Ginzburg parameter encoding the transverse correlation length of the fluid.
The dielectric susceptibility $\chi(\vec{r}-\vec{r}')$ of the system is defined as 
\begin{widetext}
\begin{equation}\label{chiparaperp}
\begin{aligned}
	\mathcal{U}_{\rm el}[\bm{\mathcal{P}}]&=\frac{1}{2\epsilon_0}\int d\vec{r} d\vec{r}'\bm{\mathcal{P}}(\vec{r}) \cdot \chi^{ -1}(\vec{r}-\vec{r}')\cdot \bm{\mathcal{P}}(\vec{r}'),\\
 &=\frac{1}{2\epsilon_0}\int d\vec{r} d\vec{r}'\left(\bm{\mathcal{P}}_\parallel(\vec{r}) \cdot\chi_\parallel^{ -1}(\vec{r}-\vec{r}')\cdot \bm{\mathcal{P}}_\parallel(\vec{r}')+\bm{\mathcal{P}}_\perp(\vec{r}) \cdot\chi_\perp^{ -1}(\vec{r}-\vec{r}')\cdot \bm{\mathcal{P}}_\perp(\vec{r}')\right),
 \end{aligned}
\end{equation}
\end{widetext}
where we have split the polarization field into a longitudinal part $\bm{\mathcal{P}}_\parallel$ and a transverse part $\bm{\mathcal{P}}_\perp$, which respectively satisfy
$\bm{\nabla}_{\vec{r}}\times  \bm{\mathcal{P}}_\parallel(\vec{r})=0$ and $\bm{\nabla}_{\vec{r}} \cdot  \bm{\mathcal{P}}_\perp(\vec{r})=0$.
The susceptibility can also be 
decomposed into longitudinal and transverse components (see Eq.~\ref{chiMFmoldyn}).
 
Inverting Eq.~(\ref{HP}), we find that the longitudinal and transverse susceptibilities are equal to 
\begin{equation}
\label{chiMF}
\chi_\parallel(q)=\frac{1}{1+K}, \quad  \chi_\perp(q)=\frac{1}{K+\kappa_tq^2}.
\end{equation}

 Figure~\ref{fig:SI} compares results for pure water obtained with molecular dynamics (MD) simulations and with the model presented in Eq.~(\ref{chiMF}). On panel (a), the longitudinal susceptibility $\chi_{\parallel}$ is plotted as a function of $q$. The longitudinal susceptibility is found to be constant and equal to the bulk susceptibility in MD simulations (blue markers). Furthermore, it is well described by Eq.~(\ref{chiMF}) (dashed blue line). Note that the permittivity of the medium  obeys $\epsilon_w=\left(1-\chi_\parallel(0)\right)^{-1}=1+1/K$. 

Panel (b)  presents the transverse susceptibility $\chi_{\perp}(q)$. The two parameters of the field theory model, namely $K$ and $\kappa_t$, were adjusted to fit MD simulations (dashed green line). The model then predicts well the behavior observed at low $q$ in MD simulations (blue markers).  
\subsection{Response function of electrolytes in the field theory framework}
The partition function for $N_+$ monovalent cations and $N_-$ monovalent anions of respective charges $e$ and $-e$ solvated in this medium can be written as
\begin{widetext}
\begin{equation}
\begin{aligned}
\mathcal{Z}&=\frac{1}{N_+!}\frac{1}{N_-!}\left[\prod_{i=1}^{N+}\int d\vec{r}_i \right]\left[\prod_{j=1}^{N-}\int d\vec{r}_j \right]\int \mathcal{D}[\bm{\mathcal{P}}]\,\exp\left[-\frac{\beta}{2\epsilon_0}\int d\vec{r} d\vec{r}' 	\bm{\mathcal{P}}(r)\cdot K(\vec{r}-\vec{r}') \cdot \bm{\mathcal{P}}(r')\right]\\
&\times \exp\left[-\frac{\beta}{2}\int d\vec{r}\int d\vec{r}'\left[\rho_i(\vec{r}) - \nabla_\vec{r}\cdot \bm{ \mathcal{P}}(\vec{r})\right]v(\vec{r}-\vec{r}')\left[\rho_i(\vec{r}') - \bm{\nabla}_{\vec{r}'}\cdot\bm{ \mathcal{P}}(\vec{r}')\right]\right]\,,
\end{aligned}
\end{equation}
where $\beta=1/(\kB T)$ and $v(\vec{r}-\vec{r}')=1/(4\pi\epsilon_0|\vec{r}-\vec{r}'|)$ denotes the Coulomb potential, while 
\begin{equation}
\label{rho}
\rho_i(\vec{r})=\sum_{i=1}^{N+} e\, \delta(\vec{r}-\vec{r}_i)-\sum_{j=1}^{N-} e\, \delta(\vec{r}-\vec{r}_j)
\end{equation}
denotes the ionic charge density.
Introducing an auxiliary field $\Phi$ and performing a Hubbard-Stratonovich transform to get rid of the long-range Coulomb potential~\cite{levy2020}, we can compute the partition function in the grand-canonical ensemble as 
\begin{equation}
	\label{GrandPartFunc}
	\Xi=\int \mathcal{D}[\bm{\mathcal{P}}]\,\mathcal{D}[\Psi]e^{ - \beta F_u[\Psi, \bm{{\mathcal{P}}}]} \,,
\end{equation}
where we have defined the action 
\begin{equation}\label{action}
\begin{aligned}
	F_u[\Psi,{\bm {\mathcal P}}]&= \frac{1}{2\epsilon_0}\int d\vec{r}d\vec{r'}\bm{\mathcal{P}}(\vec{r})\cdot K(\vec{r}-\vec{r}')\cdot \bm{\mathcal{P}}(\vec{r}')-\frac{2n}{\beta}\int d\vec{r} \cosh(\beta e \Psi )\\&-\frac{1}{2}\int d\vec{r} \left[\epsilon_0 (\bm{\nabla}_{\vec{r}} \Psi(\vec{r})^2-2 \Psi(\vec{r}) \bm{\nabla}_{\vec{r}} \cdot \bm{\mathcal{P}}(\vec{r})\right]\,,
 \end{aligned}
\end{equation}
with $n$  the ionic density defined as $n=c\, \mathcal{N}_a$, $c$ being the electrolyte concentration and $\mathcal{N}_a$ the Avogadro number.  The mean fields ($\psi $, $\bm{P}$) minimizing the action both vanish. 

The inverse susceptibility of the medium is given by:
\begin{eqnarray}
\label{DefSusc}
\left(\begin{array}{cc} \epsilon_0\chi^G & \chi^{\rm G}_{P,\psi} \\ \chi^{\rm G}_{\psi, P} & \frac{\chi^{\rm G}_{\psi,\psi}}{\epsilon_0}
\end{array}\right)(\vec{r}_1,\vec{r}_2)= \left(\begin{array}{cc}
	\frac{\delta^2 F_u}{\delta {\bm{\mathcal{P}}}_i(\vec{r}_1) \delta{\bm{\mathcal{P}}}_{j}(\vec{r}_2)} (\psi, \bm{P})   & \frac{\delta^2 F_u}{\delta {\bm{\mathcal{P}}}_{i}(\vec{r}_1) \delta \Psi(\vec{r}_2)} (\psi, \bm{P})\\
	\frac{\delta^2 F_u}{\delta \Psi(\vec{r}_1) \delta {\bm{\mathcal{P}}}_{i}(\vec{r}_2)} (\psi, \bm{P}) & \frac{\delta^2 F_u}{\delta \Psi(\vec{r}_1) \delta \Psi(\vec{r}_2)}(\psi, \bm{P})
\end{array}\right) ^{-1}.
\end{eqnarray}
Performing the functional derivative of $F_u$ in Fourier space and inverting the matrix, we obtain 
\begin{eqnarray}
\label{chisalt}
\chi_{ij}(q)&=&\chi_\parallel(q)\frac{q_iq_j}{q^2}+\chi_\perp\left(\delta_{ij}-\frac{q_iq_j}{q^2}\right) , \quad
\chi_\parallel(q)=\frac{\epsilon_w-1}{\epsilon_w}\frac{\frac{\epsilon_w}{\lambdaD^2}+q^2}{\frac{1}{\lambdaD^2}+q^2}, \quad \chi_\perp(q)=\frac{1}{K+\kappa_t q^2}\,,
\end{eqnarray}
where $\lambdaD=\sqrt{\epsilon_0\epsilon_w/2\beta n e^2}$ is the Debye length. 
These expressions show that the longitudinal susceptibility of the medium is now a function of the salt concentration via the Debye length. Conversely, the transverse susceptibility is not affected by the presence of the salt.

\section{Approximate conformal invariance}
\label{appConformal}

\end{widetext}

In the main text, we discussed the dependence of the function $\phi_u$ on geometry by varying the dimensionless parameter $x=d/\Reff$, while keeping constant the dimensionless parameter $u$  representing the ratio of radii, yielding the four curves shown on Fig.~\ref{numerical_results}. Let us now discuss another representation of this dependence, in terms of a conformally invariant geometric parameter that we will call $y$. 
This representation is inspired by results obtained in the case of two spheres  \cite{Schoger2022}, and is still relevant in the present case of two cylinders. It shows an interesting universality property of the Casimir attraction versus geometrical dimensions.

\begin{figure}[t!]
\centering
\includegraphics[scale=0.82]{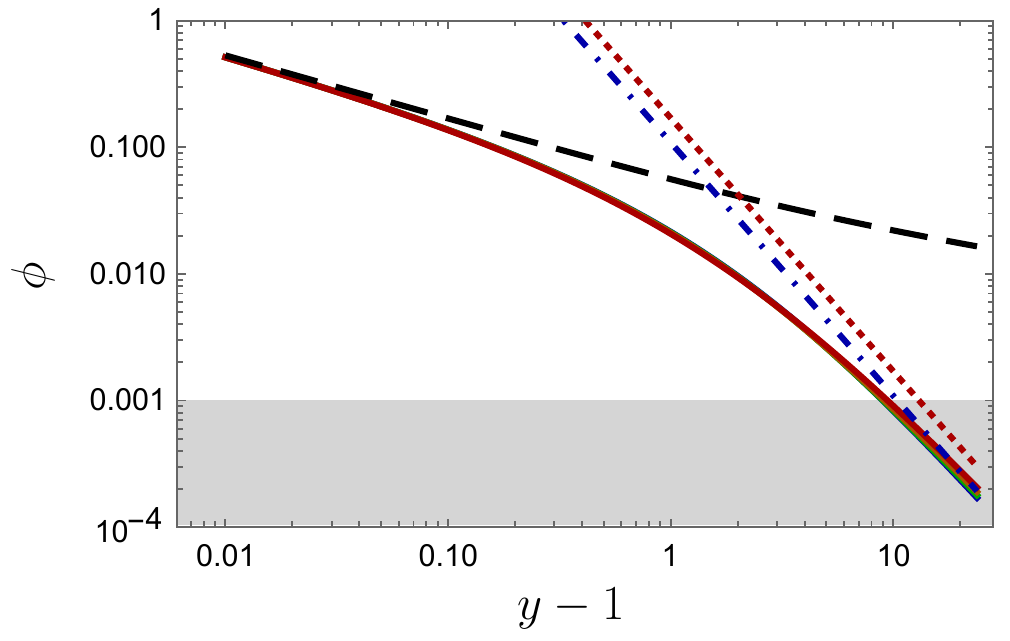}
\caption{
\textbf{Universal function $\phi$ versus the conformally invariant geometric parameter $y-1$.} Conventions and color codes are the same as in Fig.~\ref{numerical_results}. The full results (solid curves) drawn for various values of $u$ are much closer to each other in this representation than versus $x$ (Fig.~\ref{numerical_results}). They agree with the PFA expression (dashed line) at short separations and with approximations discussed in the text (dotted and dash-dotted lines) at long distances. In the grayed out band, the value of $|\caF|$  (calculated with $L/d=10^3$) is dominated by the Brownian motion energy scale $\kB T$.
}
\label{numerical_results_y}
\end{figure}

 In order to discuss this property, we first define the conformally invariant geometric parameter \cite{SchogerIJMPA2022}
\begin{equation}
y=\frac{\left(d+R_1+R_2\right)^2-R_1^2-R_2^2}{2\,R_1~R_2}~.
\label{eq:defy}
\end{equation}
In Fig.~\ref{numerical_results_y}, we show the same curves representing $\phi_u$ for four values of $u$ as in Fig.~\ref{numerical_results}, but versus $y-1$ instead of $x$.
We observe that these four curves are much closer to each other when drawn versus $y-1$ rather than versus $x$. In fact $y-1$ is a stretched version of the parameter $x$, with a stretching factor depending on $x$ and $u$:
\begin{equation}
y-1=x\left(1+\frac{u\,x}{2} \right) ~, 
\label{eq:defym1}
\end{equation}
and this stretching makes the four curves almost indistinguishable in
Fig.~\ref{numerical_results_y}. 
There remains however a small residual dependence of $\phi_u(y-1)$ on $u$, which is discussed below. 

The PFA expression \eqref{PFAres} can be written as a function of $y$ only 
\begin{equation}
\label{pfa-y}
\phi_{\rm PFA}(y)\approx \frac{H}{24}\,\sqrt{\frac2{y-1}}\,,\quad y-1\ll1~. 
\end{equation}
 However, the two long-distance expressions \eqref{long_distance} show a dependence on $u$ when written in terms of $y$:
\begin{equation}
\label{long_distance-y}
\begin{aligned}
&\phi_u(y)\approx \frac{891\pi}{16\,384(y-1)^2}\,,\quad y\gg1\,,\quad u>0~, \\
&\phi_0(y)\approx \frac{7}{64(y-1)^2}\,,\quad y\gg1\,,\quad u=0~. 
\end{aligned}
\end{equation}
These two long-distance results exhibit the same dependence on $y$, and only differ by a factor of order unity, namely ${\displaystyle 1792/(891\pi)\simeq 0.64}$. Furthermore, this long-distance result does not depend on $u$ for $u>0$.
Formulas in the first and second lines of Eq.~\eqref{long_distance-y} are shown respectively as dotted and dash-dotted lines in Fig.~\ref{numerical_results_y}. 

\begin{figure}[t!]
\centering
\includegraphics[scale=0.82]{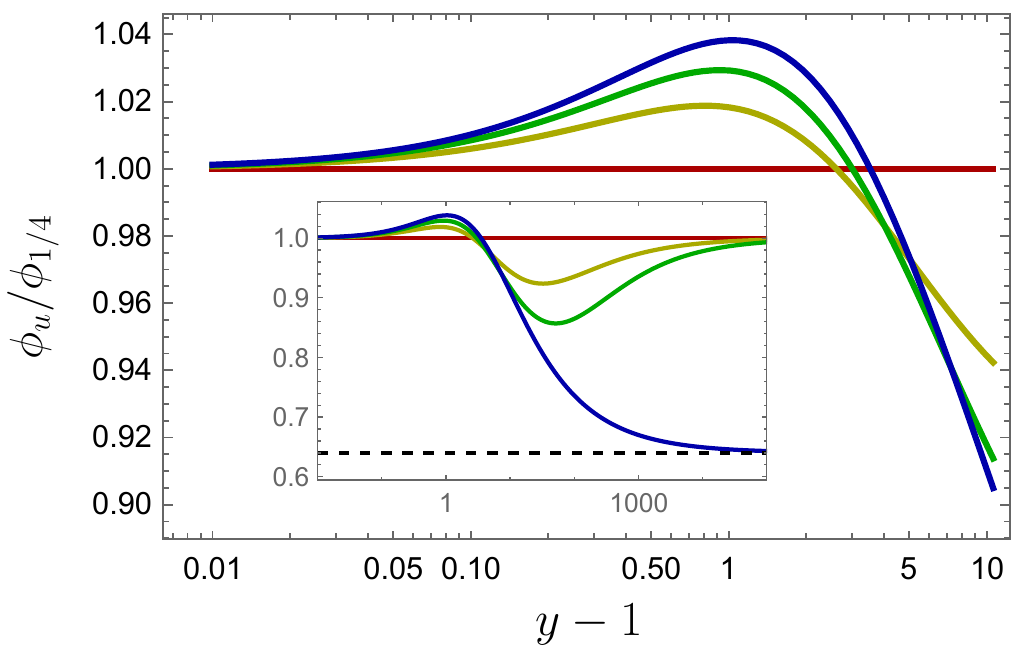}
\caption{\textbf{Ratio of the function $\phi_u$ for a given aspect ratio $u$ to the reference function $\phi_{1/4}$ corresponding to identical cylinders ($u=1/4$).} Curves are drawn versus the conformally invariant distance $y-1$. The main figure shows the domain where the Casimir interaction is larger than the thermal scale $\kB T$ (see Fig.~\ref{numerical_results_y}). The inset presents a broader distance range (same axes as in the main plot, but extended abscissa range). It shows that all curves corresponding to two cylinders with finite radii have the same asymptotic values at long distances, whereas the curve corresponding to the plane-cylinder geometry has an asymptotic value differing by a factor $\simeq 0.64$, as discussed in the text (horizontal dashed line). [Colors are the same as in Figs.~\ref{numerical_results} and \ref{numerical_results_y}.]} 
\label{ratio_zoom}
\end{figure}

More details and references regarding the significance of this approximate conformal invariance~\cite{Eisenriegler1995} are given for the case of spheres in \cite{SchogerIJMPA2022}.
Figure~\ref{numerical_results_y} shows that the representation in terms of $y$ is also relevant in the case of two cylinders discussed in the present work, where conformal invariance now corresponds to the geometry of circles in the 2d plane orthogonal to the axis of cylinders. 
The residual dependence of $\phi_u(y-1)$ on $u$ is visualized in Fig.~\ref{ratio_zoom}, where the ratio of ${\displaystyle \phi_u(y-1)}$ to ${\displaystyle \phi_{1/4}(y-1)}$ is plotted versus $y-1$. Note that we take $u=1/4$ as the reference because it corresponds to the case of two identical cylinders, which is the relevant one for our application to biological filaments (see main text).

The main plot shows this ratio for the four values of $u$ considered in Figs.~\ref{numerical_results} and~\ref{numerical_results_y}, over the domain where the Casimir interaction is larger than $\kB T$ (the plot is drawn for $L/d=10^3$). 
The inset shows a broader range of distances, allowing us to visualize the long-distance limit $y-1\gg1$. 
All curves corresponding to two cylinders with finite radii are superimposed in both the short- and the long-distance limits. Furthermore, they remain close to each other over the whole domain of variation of $y-1$ (for instance, the relative difference never exceeds 8\% for $u=0.1$). The curve corresponding to the plane-cylinder case ($u=0$) is the most different from the reference of two identical cylinders, and tends to ${\displaystyle 1792/(891\pi)}\simeq 0.64$ in the long-distance limit, thus leading to a relative difference of 36\% in this case.

\end{document}